# Ultrafast Superradiant Scintillation from Weakly Confined CsPbBr$_3$ Nanocrystals

Matteo L. Zaffalon[1], Andrea Fratelli[1], Zhanzhao Li[2], Francesco Bruni[1], Ihor Cherniukh[3], Francesco Carulli[1], Francesco Meinardi[1], Maksym V. Kovalenko[3], Liberato Manna[2], Sergio Brovelli[1*]

[1]*Dipartimento di Scienza dei Materiali, Università degli Studi di Milano Bicocca, via R. Cozzi 55, Milano, Italy*
[2] *Istituto Italiano di Tecnologia, via Morego, Genova, Italy*
[3] *Department of Chemistry and Applied Bioscience, ETH Zürich, Zürich, Switzerland.*
*Laboratory for Thin Films and Photovoltaics and Laboratory for Transport at Nanoscale Interfaces, Empa – Swiss Federal Laboratories for Materials Science and Technology, Dübendorf, Switzerland.*

**Efficiency and emission rate are two traditionally conflicting parameters in radiation detection, and achieving their simultaneous maximization could significantly advance ultrafast time-of-flight (ToF) technologies. In this study, we demonstrate that this goal is attainable by harnessing the giant oscillator strength (GOS) inherent to weakly confined perovskite nanocrystals, which enables superradiant scintillation under mildly cryogenic conditions that align seamlessly with ToF technologies. We show that the radiative acceleration due to GOS encompasses both single and multiple exciton dynamics arising from ionizing interactions, further enhanced by suppressed non-radiative losses and Auger recombination at 80 K. The outcome is ultrafast scintillation with 420 ps lifetime and light yield of ~10'000 photons/MeV for diluted NC solutions, all without non-radiative losses. Temperature-dependent light-guiding experiments on test-bed nanocomposite scintillators finally indicate that the light-transport capability remains unaffected by the accumulation of band-edge oscillator strength due to GOS. These findings suggest a promising pathway toward developing ultrafast nanotech scintillators with optimized light output and timing performance.**

The development of highly efficient and ultrafast scintillators is at the forefront of radiation detection research[1], particularly for high-precision positron emission tomography (ToF-PET) scanners in oncology[2], and for four-dimensional tracking ($x,y,z,t$) and five-dimensional calorimetry ($x,y,z,t,E$) of nuclear events in high-brightness hadron colliders[3,4]. Fast scintillation kinetics enable high time resolution and the discrimination of nearby events by reducing pileup. Simultaneously, a high light yield ($LY$), defined as the number of photons emitted per unit of deposited energy, is necessary to ensure a statistically significant event estimate against noise[5]. To meet these demands, a scintillator must combine a strong interaction with ionizing radiation, facilitated by a high average atomic number ($Z_{AV}$), minimized non-radiative losses, a large transition oscillator strength, and sufficient radiation stability (also referred to as radiation hardness) to sustain the strong irradiation levels in high energy physics contexts[6]. Unfortunately, conventional inorganic or plastic scintillators alone are increasingly inadequate. Inorganic scintillators have slow lifetimes (typically tens of nanoseconds or longer)[7,8], while plastic scintillators have insufficient $Z_{AV}$ and low radiation hardness[9,10]. To circumvent these limitations for ToF-PET, so-called meta-scintillator schemes are being explored[11], which leverage energy-sharing processes from heavy inorganic materials (such as lutetium-yttrium oxyorthosilicates or bismuth germanates) to molecular emitters with decay times of a few nanoseconds[12-16].

A promising materials-based approach with potential groundbreaking impact in all applications of ultrafast radiation detection involves the use of colloidal nanocrystals (NCs) of direct bandgap semiconductors[17-23], particularly lead halide perovskites[24-39], as nanoscintillators. These NCs offer affordable synthetic scalability and fast emission of molecular dyes combined with the high $Z_{AV}$ and radiation stability of inorganic materials[29], effectively merging the advantages of both existing material systems. Crucially, the quantum confinement regime characteristic of NCs further introduces unique nanoscale physical properties, such as size tunability of the emission wavelength [24,40,41] and large exciton and biexciton binding energies[42-45] that do not occur in bulk or molecular materials, which push the optical and timing capabilities for ToF scintillation to unprecedented levels. A notable example is the recently demonstrated ultrafast scintillation, occurring in hundreds of picoseconds, achieved through the decay of multi-excitons formed upon the interaction of NCs with ionizing radiation[19,35,36,38,46,47], which significantly surpasses the monomolecular decay observed in organic dyes.

Despite their potential, the study of NCs for ultrafast scintillation is still in its early stages, and powerful photophysical motifs demonstrated under conventional (non-ionizing) optical pumping have yet to be explored in the context of ionizing excitation. One compelling process that fully radiatively accelerates emission, without activating competing channels, such as concentration quenching or nonradiative Auger recombination – that is the nonradiative annihilation of one exciton in favour of a third carrier[42] – that affects the multi-exciton regime, is the so-called giant oscillator strength (GOS)[48-50]. This results from the coherent coupling of transition dipoles that sustain exciton delocalization over multiple unit cells, which can give rise to collective ultrafast superradiant emission, a phenomenology that has been observed in epitaxial quantum wells[49,51], molecular solids and aggregates[52] and in quantum-confined nanophases in inorganic matrixes[53].

The formation of giant transition dipoles further underlies the reduction of the gain threshold[54,55] and the radiative acceleration of luminescence in colloidal chalcogenide nanoplatelets[22,56,57] and weakly confined $CsPbBr_3$ NCs with decreasing temperature. GOS has recently been shown to produce accelerated emission from both individual particles[58] and NC superlattices[59-65], which is highly promising for quantum light sources[66]. Such capability of individual $CsPbBr_3$ NCs is enabled by their unique, compared to conventional NCs, prevalence of the radiative recombination through the so-called bright triplet state at cryogenic temperatures[67]. Suppression of thermal quenching and phonon scattering at mild cryogenic temperatures is thus accompanied by the thermalization of excitation in the bright excitonic triplet substate resulting in a massive acceleration of temporal kinetics at near-unity emission yield (see scheme in **Figure 1a**)[58]. This is invaluable for the design of ultrafast emitters because it breaks the common dichotomy between luminescence efficiency and decay rate, which are typically anticorrelated in conventional scintillators, where acceleration of emission kinetics is achieved by activating parasitic non-radiative pathways that quench the efficiency. Also, crucially for scintillation, a large particle size enhances the energy deposition capability of NCs following interaction with ionizing radiation, promoting the formation of fast-emitting high order multiexcitons[68-70], which are less affected by nonradiative Auger recombination owing to the universal scaling law of the Auger time constant with the particle volume ($\tau_{AR} \propto V$)[71,72]. Importantly, as we show here, the mid-range translational order in large NCs also leads to a temperature dependence of the Auger process imposed by the conservation of energy and momentum laws (typical of bulk solids), resulting in the complete suppression of Auger losses at low temperatures. Overall, this makes multi-exciton scintillation more efficient than in smaller particles even at room temperature and pushes the multi-exciton yield to unit values in mild cryogenic conditions due to the complete suppression of Auger losses. The combination of these unique properties of weakly confined CsPbBr$_3$ NCs, namely, single particle GOS, high interaction capability with ionizing radiation, and suppressed Auger decay, offers an unprecedented opportunity for the desired simultaneous optimization of timing and efficiency performance in a new class of ultrafast superradiant nanoscintillators, a field that has never been explored.

Here, we demonstrate the potential of GOS in weakly confined CsPbBr$_3$ NCs for ultrafast radiation detection by studying the photophysics and scintillation down to liquid nitrogen temperature, a mild cryogenic condition readily accessible for medical instrumentation and typical of many high-energy physics experiments. Photoluminescence (PL) measurements as a function of temperature on CsPbBr$_3$ NCs with lateral size of 15.9 nm confirm the occurrence of single particle GOS, resulting in 20-fold acceleration of the luminescence lifetime at 80 K accompanied by the saturation of the PL quantum yield ($\Phi_{PL}$) at near unity values. This behaviour is remarkably followed also by biexcitons, probed here via time-resolved PL and transient absorption (TA) experiments as a function of temperature and excitation fluence, showing a gradual radiative decrease of the biexciton lifetime accompanied by the complete suppression of the nonradiative Auger recombination, leading to accelerated biexciton dynamics with nearly unity efficiency at 80 K. These behaviours are completely preserved under ionizing excitation resulting in fully radiative ultrafast scintillation with 420 ps decay time and *LY* as high as 10'000 photons/MeV at 80 K for a diluted NC solution which is highly promising for ToF radiation detection technologies. Finally, light propagation measurements of the luminescence of testbed nanocomposites containing the same large CsPbBr$_3$ NCs, corroborated by Monte Carlo ray tracing simulations, indicate that the temperature-induced spectral modifications (narrowing and red shift) do not affect light transport, resulting in a net positive enhancement of the scintillation output and preserved ultrafast dynamics. These results, therefore, suggest a potential strategy for the future development of high-performance nanotechnological scintillators for ultrafast radiation detection.

**Results and Discussion**

CsPbBr$_3$ NCs were synthesised by a modified hot injection procedure described in the Methods[73] followed by ligand exchange with dodecyl ammonium bromide. Structural analysis by high-resolution transmission electron microscopy (HR-TEM) and X-ray diffraction (XRD) in **Figure 1b** and **Supporting Figure S1** shows that the NCs have a cuboid shape with an average size of 15.9±1.4 nm, well above the Bohr diameter (~7 nm) of CsPbBr$_3$[41], and orthorhombic crystal structure. In order to investigate the optical and scintillation properties of isolated NCs without additional effects due to interparticle interactions and/or mutual cross-excitation effects via released photoelectrons, the NCs were dispersed in a non-scintillating solvent (octane) at relatively low concentration (0.7wt.%) and tested in this form, except where explicitly noted. The room temperature optical absorption spectrum and PL profiles show an absorption edge at ~2.42 eV and PL peak at 2.38 eV (**Figure 1c**), close to the emission energy of bulk CsPbBr$_3$ (~2.38 eV)[74]. The room temperature PL dynamics (**Figure 1d**) measured at vanishingly low excitation fluence to ensure the production of only single excitons (average NC exciton occupancy ⟨N⟩~0.05) is essentially single exponential with an effective lifetime (defined as the time after which the signal has dropped by a factor equal to $e$) of $\tau_X^{PL}(300K)$~21 ns (corresponding to a single exciton decay rate, $k_X(300K) = 48$ MHz), measurably longer than smaller CsPbBr$_3$ NCs due to the stronger $s$-$p$ hybridization in larger particles[75]. Note that we consider the single exciton lifetime as the radiative lifetime, since non-radiative losses in NCs are typically due to ultrafast trapping processes that occur prior to exciton recombination.

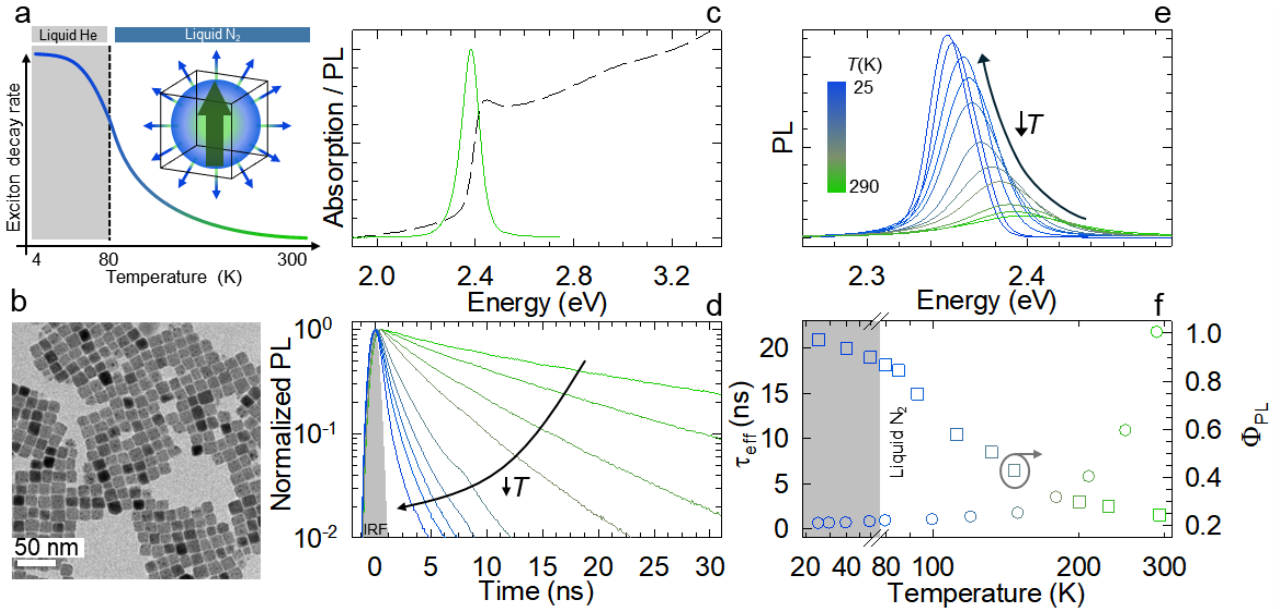

**Figure 1: Structural and $T$-dependent optical properties of single excitons a)** Schematic of the effect of GOS on the decay kinetics of in weakly confined CsPbBr$_3$ NCs (typically with lateral size ≥9 nm or so) with decreasing temperature. **b)** High-resolution TEM micrographs of CsPbBr$_3$ NCs. **c)** Normalized optical absorption (dashed line) and PL (solid line) spectra of CsPbBr$_3$ NCs in toluene. **d)** Normalized $T$-dependent PL decay traces of CsPbBr$_3$ NCs excited using ps-pulsed 3.06 eV laser operated at 1 MHz. **e)** $T$-controlled PL spectra of CsPbBr$_3$ NCs in the 290-25 K range when excited at 3.06 eV. **f)** PL efficiency ($\Phi_{PL}$) values (squares) obtained rescaling the PL efficiency measured at room temperature for the integrated area of $T$-dependant PL spectra from 'e'. The circles correspond to the measured effective PL decay time from 'd'.

To promote the formation of GOS, we cooled down the NC solution and collected both the steady-state and time-resolved PL still carefully adjusting the excitation fluence to operate in the single exciton regime. The PL spectra in the 290-25 K temperature range are reported in **Figure 1e** showing an almost 4-fold intensification upon cooling together with the spectral narrowing and red-shifting characteristic of perovskite NCs. As a result, the corresponding PL efficiency in **Figure 1f** increases from $\Phi_{PL}$~21% (as measured with an integrating sphere) at room temperature up to nearly 90% at 80 K due to suppression of trapping (extracted from the raise of integrated PL intensity vs. $T$). In contrast to conventional observations of slower decay dynamics at low temperatures and in agreement with previous reports[58,76-78], cooling the NCs led to over twenty-fold acceleration of the single exciton lifetime to $\tau_X^{PL}(80K) = 830$ ps (**Figure 1d** and **1f**). Notably, the PL efficiency is essentially maximised at liquid N$_2$ temperature, with marginal, <10%, improvement below 30 K when the $\Phi_{PL}$ reached ~97%, which corroborates the technological validity of the approach.

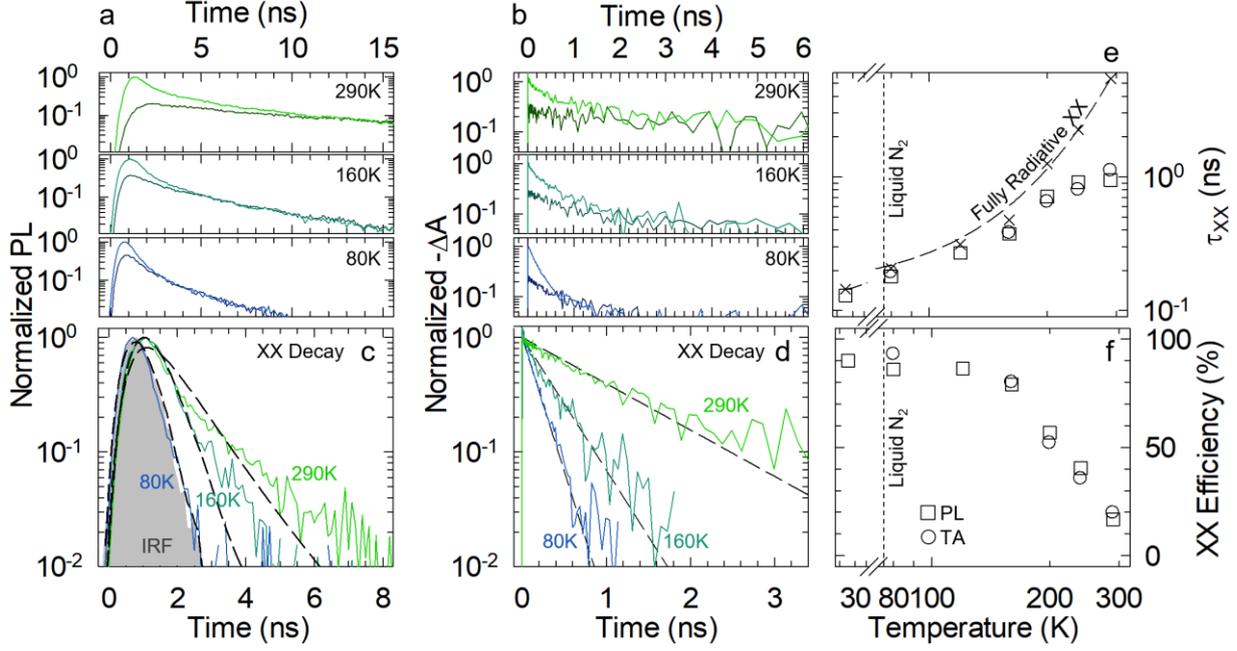

**Figure 2:** *T*-dependant biexciton efficiency and Auger processes **a)** PL decay traces of CsPbBr$_3$ NCs collected at controlled temperatures (290K, 160K, and 80K) excited with pulsed 3.06 eV fs-laser at low (dark coloured) and high (light coloured) fluences. **b)** Transient absorption bleach recovery dynamics collected at the same temperature in 'a' under similar low- and high-fluence regimes. **c)** PL decay traces of the biexcitonic components extracted from 'a'. The grey shaded area represents the instrument response (IRF), and the dashed lines correspond to the best fit with a single-exponential decay function convoluted with the IRF. **d)** Differential TA curves of the biexciton components extracted from 'b' together with the best single exponential decay fit (dashed lines). **e)** Temperature dependence of the biexciton lifetime ($\tau_{XX}$) extracted from fluence dependent PL (squares) and TA (circles) measurements. The crosses correspond to the theoretical fully radiative biexciton lifetime calculated as $\tau_x(T)/4$. The dashed line is a guide for the eye. **f)** Calculated biexciton efficiency as a function of temperature from both set of PL and TA measurements. The same colour scheme applies to the figure.

Having confirmed the necessary photophysical prerequisites of weakly confined CsPbBr$_3$ NCs in the single-excitonic regime, before moving to the study of scintillation, we assessed the effect of temperature on the multi-excitonic recombination that underpins the scintillation process. To this end, we performed PL and TA measurements as a function of temperature with increasing excitation fluence. **Figures 2a** and **2b** show the low and high fluence PL and TA bleaching dynamics, respectively, at three representative temperatures. Consistent with the PL data in **Figure 1d**, in both experiments, lowering the temperature accelerates single exciton decay while increasing the excitation fluence leads to the rise of a fast component due to biexciton recombination. Following the procedure introduced by Klimov[79] for the analysis of multi-exciton dynamics, **Figure 2c** and **2d** show the biexciton decay curves extracted by subtracting the single-exciton dynamics from the respective high fluence curve after tail normalisation at long delay times. In agreement with previous reports[80], both experiments yield a room temperature biexciton lifetime $\tau_{XX}(300K)$~1 ns (corresponding to a biexciton rate $k_{XX}(300K) =1$ ns$^{-1}$). Considering the statistical link between the single and biexciton radiative rates ($k_{XX}^{rad}=4k_X$), this corresponds to a biexciton efficiency $\Phi_{XX}(300K) = k_{XX}^{rad}(300K)/k_{XX}(300K) = 19\%$. We emphasize that because of the temperature dependence of both GOS, the population equilibrium between the excitonic substates and the Auger recombination in weakly confined NCs, all decay rates are functions of temperature. This is not the case in strongly confined particles where the Auger recombination is temperature independent due to the breakdown on momentum conservation requirement[72,81]. As a result, following the decrease of $\tau_X(T)$ with decreasing temperature, the biexciton decay also gradually accelerates, as further quantified by the corresponding experimental $\tau_{XX}(T)$ in **Figure 2e**, which approaches the fully radiative value at ~100 K, indicating essentially complete suppression of Auger recombination. This important effect of temperature on biexciton photophysics is further highlighted in **Figure 2f**, where we report the $\Phi_{XX}$ vs. $T$ trend showing saturation at >90% at liquid N$_2$ temperatures. Crucially, similar to the single-exciton behaviour discussed in **Figure 1**, this increase in biexcitonic recombination efficiency is accompanied by its full radiative acceleration, effectively extending the efficiency-velocity paradigm to multi-carrier dynamics. We note that the time dynamics of biexcitons as a function of temperature in **Figure 2** suggests that biexcitons in our NCs do not experience any measurable additional GOS over their respective single excitons. This differs from the observation of biexciton GOS in bulk halides at low temperatures[82], which exceeds the oscillator strength of single excitons by orders of magnitude due to the much larger size of biexcitons, effectively involving a massively larger number of unit cells, and is likely due to the size limit imposed on both single and biexcitons by quantum confinement.

Based on the promising optical properties of our CsPbBr$_3$ NCs in both the single and multi-exciton regime, we proceeded to demonstrate their potential in scintillation using X-ray excitation. The RL spectrum of a colloidal suspensions of NCs in octane (0.7 wt%) is shown in **Figure 3a** and closely matches its corresponding PL profile indicating that the RL originates from the same band-edge exciton states. A photograph of the sample under X-ray irradiation in shown in **Figure 3b**. The $LY$ of 150 µL (in a 0.5 cm high cylindrical cuvette) of the colloidal suspension measured in the same excitation and collection geometry at room temperature as a commercial plastic scintillator EJ-276D ($LY$ = 8600 photons/MeV) of the same volume (in both cases resulting in essentially complete deposition of the incident X-ray excitation) is 2400±100 photons/MeV, consistent with previous results on diluted CsPbBr$_3$ NCs composites[34]. We emphasize that the definition of $LY$ for colloidal NCs is still a matter of debate as it depends on both single particle effects (i.e. stopping power, exciton and biexciton emission yields) and extensive phenomena such as outcoupling and interparticle cross-excitation by photoelectrons released outside a single NC following primary excitation events, which can be significantly improved by increasing the particle concentration and packing, as recently demonstrated using dense CsPbBr$_3$ NCs-based composites[38,39]. Since the scope of this work is to demonstrate the potential of intraparticle GOS, the sample concentration was specifically chosen to avoid interparticle effects. Consistent with the PL trend, the RL peak at 2.38 eV red-shifts, narrows and intensifies by over four-fold upon cooling (**Figure 3c, 3d**), reaching a $LY(80K)$ ~ 10'000 photons/MeV, corresponding to the radiative limit of the scintillation efficiency for this system.

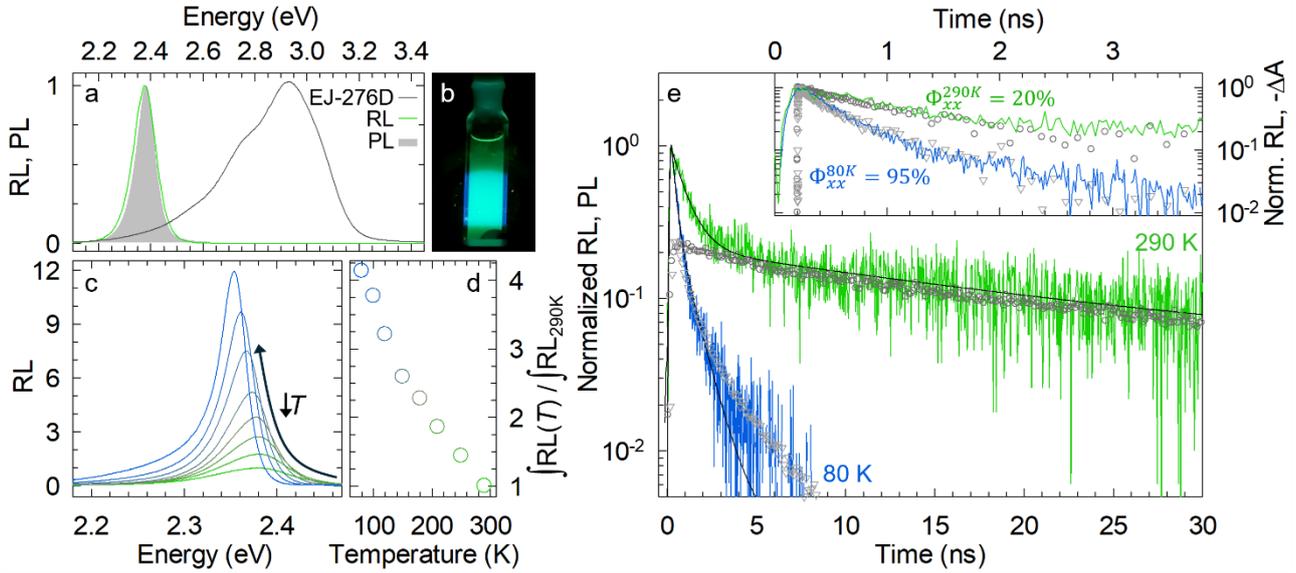

**Figure 3: GOS-enhanced scintillation of CsPbBr$_3$ NCs. a)** RL spectrum of a solution of CsPbBr$_3$ NCs dispersed in octane (0.7 wt%) together with a commercial EJ-276D plastic scintillator used as reference. The corresponding normalized PL spectrum of CsPbBr$_3$ NCs is shown as a shaded grey area. **b)** Picture of an octane colloidal suspension of CsPbBr$_3$ NCs under X-ray excitation. **c)** X-ray excited (~7 keV) RL spectra of CsPbBr$_3$ NCs as a function of temperature in the range 290-80 K. **d)** Temperature dependence of the spectrally integrated RL intensity extracted from 'c'. The values have been normalized for the value at 290 K. **e)** Time-resolved PL (symbols) and RL (coloured lines) decay traces of CsPbBr$_3$ NCs normalised over the single-excitonic tail, both at 290 K (circles, green line) and at 80 K (triangles, blue line). The solid black lines are the best fit of the scintillation decays (see Supporting Information for details). Inset: comparison of the scintillation dynamics and the bleach recovery dynamics of biexciton components (symbols) extracted from TA measurements at 290 and 80 K. The same colour scheme applies to the inset.

Also importantly for ToF applications, the RL kinetics undergoes a similar acceleration as the respective PL. Specifically, the room temperature RL decay trace (**Figure 3e**) features two contributions typical of CsPbBr$_3$ NCs[34,35,38,46]: a main component with relative weight $w_X(300K) = 80\%$ and lifetime $\tau_{RL}^X(300K) = 20$ ns which coincides with the low-fluence PL decay (circles in **Figure 3e**) and is attributed to the RL of single excitons, and a faster contribution with $w_{XX}(300K) = 20\%$ and lifetime $\tau_{RL}^{XX}(300K) = 850$ ps closely matching the corresponding biexciton decay extracted from the TA dynamics (symbols in the inset of **Figure 3e**) and thus attributed to biexciton emission with efficiency of ~20%. Consistent with recent results, the ratio between the amplitudes of the $X$ and $XX$ components, following the typical approach for ultrafast kinetic studies in NCs, yields an estimated excitonic population $\langle N \rangle = 3.5$ as expected for ~16 nm NCs[79,80].

At 80 K, both the single and biexciton decays are significantly accelerated with $\tau_{RL}^X(80K) = 920$ ps and $\tau_{RL}^{XX}(80K) = 220$ ps, yielding a biexciton efficiency as high as 95% also under X-ray excitation. As a result of the faster dynamics in both excitonic regimes, the so-called effective scintillation lifetime, calculated as the weighted harmonic average of the single and biexciton and contributions ($(\tau_{RL}^{EFF})^{-1} = \sum \frac{w_i}{\tau_i}$), is as fast as $\tau_{RL}^{EFF}(80K) = 420$ ps, which is very promising for fast timing technologies. The potential advantage of the peculiar fully-radiative ultrafast photophysics of weakly confined CsPbBr$_3$ NCs becomes particularly evident by estimating the coincidence time resolution[47] (*CTR*) ideally achievable in ToF-PET scanners at mild cryogenic temperatures. This empirical parameter expresses the time required to obtain a statistically relevant signal given the effective time and the *LY* and, although not formalized identically, is also a representative figure of merit for application in ultrafast calorimeters for high luminosity colliders, which require double-impulse separations of a few nanoseconds in addition to time resolutions of a few tens of ps or less. Using the expression $CTR = 3.33\sqrt{\frac{\tau_{RISE} \times \tau_{RL}^{EFF}(80K)}{\Gamma(80K)}}$, where $\tau_{RISE}$ is the 10-90% signal rise time here set to 90 ps and $\Gamma(80K) = 5100$ is the estimated number of scintillation photons emitted at 80 K under the characteristic 511 keV gamma excitation used in ToF-PET, we obtain an estimated $CTR(80K) = 9$ ps, which is very promising for meeting the so-called 10 ps challenge[83] that would allow millimetre spatial resolution in ToF-PET diagnostics.

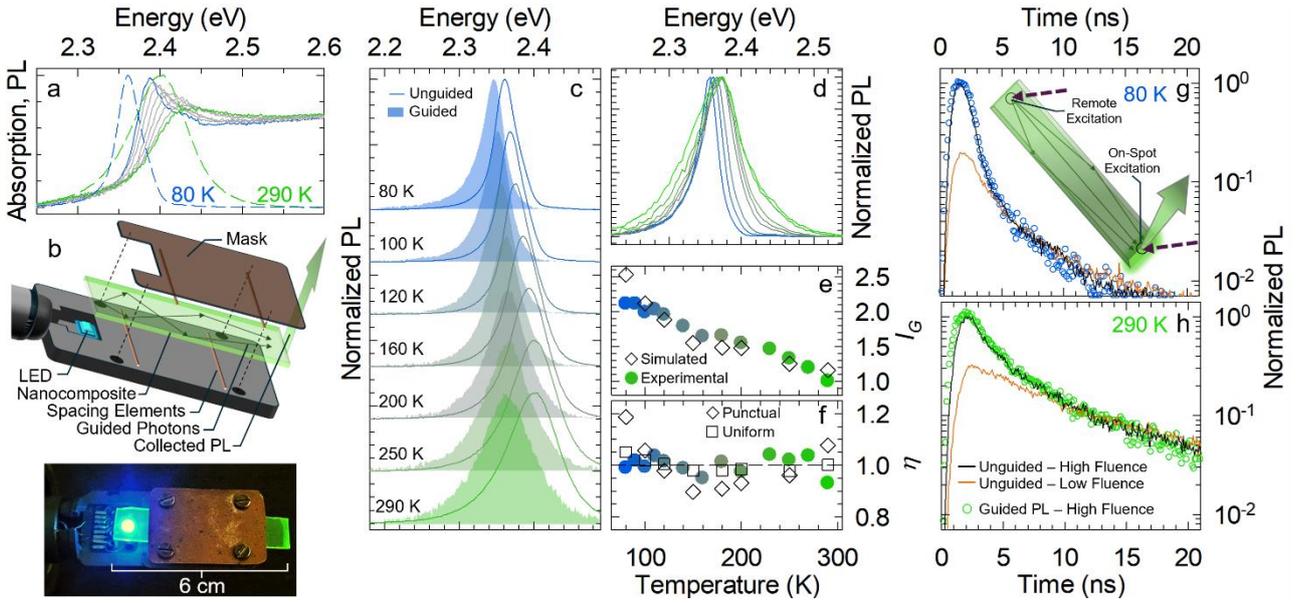

**Figure 4: Temperature effects on light transport. a)** Optical absorption spectra at decreasing temperature from 290 K (green line) to 80 K (blue line). The corresponding normalized PL spectra at the highest and lowest temperatures are shown with dashed lines. **b)** Schematic representation and photograph of the experimental configuration used to collect PL spectra shown in 'c'. **c)** Normalized PL spectra (solid lines) at controlled temperatures together with the corresponding guided PL spectra (shaded areas) excited at ~6 cm from the collection edge. **d)** Simulated guided PL spectra as a function of temperature. The colour scheme of 'c' applies. **e)** Spectrally integrated intensity of the guided PL as a function of temperature. Data were normalized to the value at 290 K. Circles and diamonds represent the experimental and the simulated data, respectively. **f)** Guided PL emission efficiency (η) as a function of temperature (filled circles for the experimental data) together with corresponding simulated values (empty diamonds). The simulated η-values under uniform illumination of the waveguide are shown as empty squares. Tail normalized guided (circles) and unguided (lines) PL decay traces at (**g**) 290 K or (**h**) 80 K under low (orange line) and high (black line) excitation fluences.

Finally, given the spectral change in absorption and emission profiles at cryogenic temperatures (**Figure 4a**) associated with the decrease in thermal disorder and the consequent accumulation of oscillator strength in band-edge states, it is instructive to evaluate the effect of temperature on the light guiding ability of scintillating waveguides based on weakly confined NCs. This is because, while lowering the temperature reduces the Stokes shift between emission and absorption[84], suggesting more reabsorption losses, the concomitant suppression of homogeneous broadening might reduce the spectral overlap sufficiently to lead to a net improvement in light-guiding ability. To assess this aspect, we fabricated testbed nanocomposite waveguides (6 x 0.5 x 0.1 cm) by optical polymerization of poly-lauryl methacrylate containing NCs and evaluated the effect of temperature on light propagation according to the experimental scheme shown in **Figure 4b**. The polymer matrix was chosen to be non-scintillating in order to study the response of the NCs alone, and the loading was set at 0.1 wt % to keep light scattering low despite the large particle size and the absence of resurfacing steps with compatibilizing ligands. First, the nanocomposites were optically characterized to ensure the preservation of the optical properties of the NCs after radical polymerization.

The PL spectrum and temporal dynamics are shown in Supplementary **Figure S2** and demonstrate the substantial stability of the particles with respect to the fabrication process, with a minimal $\Phi_{PL}$ loss of ~5%. The nanocomposites were then placed in the variable temperature insert of a liquid helium cryostat, which allows the sample to be uniformly cooled without thermal contact with the substrate, thus preserving the waveguiding capability. The experiment then consisted of exciting one end of the nanocomposite with a 465 nm diode (inserted directly into the cryostat, see **Figure 4b**) and measuring the luminescence emitted from the opposite end (guided, $G$) as a function of temperature. This configuration represents the most unfavourable situation, as the generated light must travel the entire length of the sample before being outcoupled and was specifically chosen to make the effect of temperature on reabsorption more apparent. The luminescence at the point of excitation (unguided, $UG$) was also collected to directly estimate the evolution of the spectral shape and intensity with temperature, which were then used to model the light transport via Monte Carlo simulations. **Figure 4c** shows the normalized spectra of guided and unguided PL for temperatures ranging from 300 K to 80 K, the corresponding simulated guided spectra are shown in **Figure 4d**. A systematic redshift of the guided spectrum with respect to the corresponding unguided emission is observed due to partial reabsorption and re-emission by the NCs. Importantly, the intensity of guided luminescence ($I_G$) gradually increases with decreasing temperature (**Figure 4e**). To disentangle the effects of light transport and increasing $\Phi_{PL}$ with decreasing temperature, we therefore calculated the propagated emission efficiency, $\eta=I_G/I_{UG}$, expressed by the ratio between the integrated intensity of the guided and unguided light. As shown in **Figure 4f** (filled symbols), $\eta$ is found to be temperature independent, indicating that the activation of GOS does not come at the expense of the optical performance of the composite. Ray tracing Monte Carlo simulations of light propagation performed using the experimental sample geometry and optical parameters (absorption and PL spectra, $\Phi_{PL}$) show very good agreement with the experimental data (empty diamonds in **Figure 4f**). Also importantly, similar side-by-side PL measurements performed using pulsed UV laser excitation show that the guided PL retains the same ultrafast multiexciton emission dynamics as in unguided mode both at room temperature and at 80K (**Figure 4g**). Finally, the agreement between the experimental and theoretical results further allows us to use the Monte Carlo simulation to model the expected efficiency evolution with temperature under homogeneous illumination of the waveguide body (empty squares, **Figure 4f**), which might be a condition closer to the real use of scintillator materials in radiation detection (especially in sampling calorimeters). As highlighted by the dashed line in **Figure 3f** this condition indeed leads to completely temperature independent light guiding.

In conclusion, we have shown that weakly confined $CsPbBr_3$ NCs exhibit single particle GOS even under ionising excitation, resulting in fully radiatively accelerated single and biexciton dynamics at mild cryogenic temperatures. This is accompanied by enhanced single and biexciton emission due to suppressed thermal losses and non-radiative Auger recombination, effectively allowing the radiative limit of both kinetics and efficiency to be reached without sacrificing light transport capability. Furthermore, we anticipate that excellent margins for further improvement are in principle possible by exploiting the ability of perovskite nanocrystals to self-assemble into dense superlattices[85], which exhibit further superradiance at the supramolecular level[60,61,64,65] and realistically offer the possibility of increased energy capture upon ionizing excitation and consequently even higher light yields. Finally, their solution processability makes them fully compatible with metascintillator and/or cavity-based designs, which further enhance the timing capability by increasing the LY and accelerating the radiation decay via the Purcell effect[86]. The superradiant scintillation motif thus opens up interesting possibilities for the future development of ultrafast nanotechnological scintillators for fast time-of-flight radiation detection.

## Acknowledgments

The authors wish to thank Alexander L. Efros (Naval Research Laboratory) for his valuable insights into giant oscillator strength in confined multiexcitons. This work was funded by Horizon Europe EIC Pathfinder program through project 101098649 – UNICORN, by the PRIN program of the Italian Ministry of University and Research (IRONSIDE project), by the European Union—NextGenerationEU through the Italian Ministry of University and Research under PNRR—M4C2-I1.3 Project PE_00000019 "HEAL ITALIA", by European Union's Horizon 2020 Research and Innovation programme under Grant Agreement No 101004761 (AIDAINNOVA). This research is funded and supervised by the Italian Space Agency (Agenzia Spaziale Italiana, ASI) in the framework of the Research Day "Giornate della Ricerca Spaziale" initiative through the contract ASI N. 2023-4-U.0t

## Competing interests

The authors declare no competing interests.

# Ultrafast Superradiant Scintillation from Weakly Confined CsPbBr$_3$ Nanocrystals

Matteo L. Zaffalon[1], Andrea Fratelli[1], Zhanzhao Li[2], Francesco Bruni[1], Ihor Cherniukh[3], Francesco Carulli[1], Francesco Meinardi[1], Maksym Kovalenko[3], Liberato Manna[2], Sergio Brovelli[1*]

[1]*Dipartimento di Scienza dei Materiali, Università degli Studi di Milano Bicocca, via R. Cozzi 55, Milano, Italy*
[2] *Istituto Italiano di Tecnologia, via Morego, Genova, Italy*
[3] *Department of Chemistry and Applied Bioscience, ETH Zürich, Zürich, Switzerland.*
*Laboratory for Thin Films and Photovoltaics and Laboratory for Transport at Nanoscale Interfaces, Empa – Swiss Federal Laboratories for Materials Science and Technology, Dübendorf, Switzerland.*

**Materials and Methods**

*Materials*

Octadecene (90%), cesium(I) carbonate (Cs$_2$CO$_3$, 98%), lead (II) acetate trihydrate (Pb(OAc)$_2$·3H$_2$O, 99.99%), oleic acid (OA, 90%), didodecyldimethylammonium bromide (DDABr, 98%), toluene (anhydrous, 99.8%), methyl acetate (99.5%), lauryl methacrylate (LMA, 98%), ethylene glycol dimethacrylate (EGDM, 97.5%), 2,2-dimethoxy-2-phenylacetophenone (Irgacure 651, 99%) were purchased from Sigma-Aldrich. Didodecylamine (>97%) and Benzoyl bromide (98%) were purchased from Tokyo Chemical Industry (TCI). All chemicals were used without any further purification.

*Synthesis of CsPbBr$_3$ nanocubes*

CsPbBr$_3$ NCs are synthesized following a previously reported method[1] with minor adjustments. 0.05 mmol of Cs$_2$(CO$_3$), 0.2 mmol of Pb(OAc)$_2$·3H$_2$O and 1.25 mmol of didodecylamine were dissolved in 10 ml of octadecene and 1.5 ml of oleic acid in a 25 ml three-necked flask. The resulting mixture was pumped to vacuum at room temperature for 30 min and then at 100 °C for 1 hour. Then the mixture was changed to a nitrogen atmosphere, and the temperature was raised to 160°C. A benzoyl bromide solution (prepared by mixing 50 µL of benzoyl bromide (0.42 mmol) in 500 µL of degassed octadecene) was swiftly injected, and the reaction was run 2 minutes, after which the flask was swiftly cooled to room temperature by immersion in an ice-water bath. The NCs were precipitated by the addition of a mixture of methyl acetate and toluene (volume ratio of 2:1) to the crude solution, followed by centrifugation at 6000 rpm for 10 min. The precipitate was redispersed in 4 ml of anhydrous toluene.

*Ligand Exchange with DDABr*

2 mL of DDABr solution (0.2M in toluene) was added to the 4 mL of CsPbBr$_3$ NCs solution under vigorous stirring for 1 minute. Then, the NCs were washed by addition of 4 mL of methyl acetate followed by centrifugation at 6000 rpm for 10 minutes. The precipitate was redispersed in 4 mL toluene. One additional cycle of ligand exchange was carried out following the same procedure. Finally, the NCs solution was washed by the addition of 4 mL of methyl acetate followed by centrifugation and resuspension in toluene. This final washing was done twice. The final NCs was redispersed in 4 mL toluene for XRD and TEM characterizations. We noted that, without this final washing procedure, the CsPbBr$_3$ NCs treated with DDABr tend to degrade over time.

*Structural characterization*

*X-ray Diffraction (XRD).* XRD analysis was performed on a PANanalytical Empyrean X-ray diffractometer equipped with a 1.8 kW Cu Kα ceramic X-ray tube (λ = 1.5406 Å) and an operating at 45 kV and 40 mA. CsPbBr$_3$ NC solutions were first concentrated under a flow of nitrogen, then they were drop-cast on a zero-diffraction single crystal substrate.

*Transmission Electron Microscopy (TEM) Characterization.* Bright-field TEM (BF-TEM) images with a large field of view were acquired on a JEOL JEM-1400Plus microscope with a thermionic gun (LaB$_6$ crystal), operated at an acceleration voltage of 120 kV.

*Fabrication of polymer nanocomposite*

The polymerization was performed by adding 2,2-dimethoxy-2-phenylacetophenone photo-initiator (0.33 wt%) to a colloidal suspension of NCs in LMA/EGDM (80:20). The mixture was then transferred into a sealed mold consisting of two plain glass slabs separated by a silicone gasket and placed in a polymerization chamber with continuous 365 nm light exposure. After 15 minutes of UV irradiation, an optical-grade nanocomposite without macroscopic phase segregation was obtained.

*Optical characterization*

The UV-Vis absorption spectra, also measured as a function of temperature, were collected using a Lambda 950 spectrophotometer (Perkin Elmer) equipped with an integrating sphere. Photoluminescence (PL) spectra and corresponding PL quantum yield were measured under low-intensity continuous-wave (*cw*) excitation at 3.06 eV (405 nm), collecting the emitted light with an integrating sphere and a high-resolution CCD spectrometer (TM series, Hamamatsu). PL decay dynamics were recorded using a Si-phototube coupled to a Cornerstone 260 1/4 m VIS-NIR monochromator (ORIEL) and a time-correlated single-photon counting unit (TCSPC, time resolution 100 ps), while the sample was excited at 3.06 eV using a ~70-ps pulsed laser. Temperature-dependent measurements were carried out by mounting the sample in a closed-circuit He cryostat with optical access. Ultrafast transient absorption (TA) spectroscopy measurements were conducted using a Helios TA spectrometer (Ultrafast Systems). The laser source was a 10 W Hyperion amplified laser, operated at 1.875 kHz and producing ~260 fs pulses at 1030 nm. This was coupled to an independently tunable APOLLO-Y optical parametric amplifier (OPA) from the same supplier, producing excitation pulses at 3.1 eV synchronously chopped at 937 Hz. The probe beam was a white-light supercontinuum. The PL dynamics in the low- and high-fluence regimes were collected using the previously described TCSPC setup, coupled to the fs-pulsed output of the OPA to achieve high fluences. All spectroscopic and radiometric measurements were performed on colloidal dispersions of NCs in octane, a solvent chosen specifically for its lack of intrinsic scintillation to ensure that the radioluminescence signal originated solely from the NCs. Temperature-dependent light transport measurements were performed on a 0.1 wt% loaded PLMA nanocomposite placed in a liquid-He closed-circuit cryostat operated in an evaporated He atmosphere at ~10 mbar, ensuring uniform cooling with minimal contact with the sample holder to preserve the light-guiding properties. In this configuration, PL was excited using a narrow emitting 465 nm diode (VLDB1232G-08, Vishay) operated at constant current using a Keithley Model 2450 SourceMeter.

*RL measurements*

Unfiltered X-rays were produced using a Philips PW2274 X-ray tube with a tungsten target, equipped with a beryllium window and operated at 20 kV to produce a continuous X-ray spectrum through bremsstrahlung. Cryogenic RL measurements were conducted in the temperature range of 20−290 K using a closed-cycle He cryostat. The scintillation light was detected using a liquid-nitrogen-cooled, back-illuminated, UV-enhanced CCD detector (Jobin Yvon Symphony II), coupled to a monochromator (Jobin Yvon Triax 180) with a 100 lines/mm grating.

*LY measurements*

Light yield values were determined by comparing the integrated RL intensity under 20 kV X-ray excitation ($\langle E \rangle$ ~ 7 keV) with identical experimental conditions for a 0.7 wt% octane solution of $CsPbBr_3$ NCs placed in a 5 mm long crucible and a commercial EJ-276D plastic scintillator (LY = 8600 photons/MeV) of the same size and geometry used as a reference. In both cases, the sample size was chosen to ensure complete attenuation of the excitation beam.

*Time resolved scintillation measurements*

The time-resolved RL was measured using a pulsed X-ray source consisting of a 405 nm ~70-ps pulsed laser hitting the photocathode of an X-ray tube (N5084, Hamamatsu) set at 40 kV. The emitted scintillation light was collected using an FLS980 spectrometer (Edinburgh Instruments) coupled to a PicoHarp 300 hybrid photomultiplier tube operating in TCSPC mode. RL dynamics at low temperatures were collected in the same setup while keeping the sample constantly submerged in liquid nitrogen. The RL decay curves were analysed using a least-squares fitting approach with the following formula, which accounts for the convolution with the instrument response function (IRF):

$$F(t) = IRF(t) \otimes \left( H(t - t_0) \cdot \left[ \sum_{i=1}^{2} a_i \cdot e^{-t/\tau_i} \right] \right) + C$$

where $t_0$ corresponds to the start of the emission process, $C$ is the electronic background noise floor, and $H$ is the Heaviside function. The experimental IRF was well described by a Gaussian profile (FWHM = 120 ps), and the weight of each component ($w_i$) was calculated as the integral of each convoluted function over the entire time window. The average lifetime was calculated using the re-normalized ratio of all components $\tau_i$ according to:

$$\tau_{eff} = \left( \frac{\tau_1}{w_{1n}} + \frac{\tau_2}{w_{2n}} \right)^{-1}, \qquad w_{in} = \frac{w_i}{w_1 + w_2}$$

The same model was used to fit the biexciton dynamics obtained from the time-resolved PL measurements in Figure 2.

*Monte Carlo Ray-Tracing Simulation.*

Simulations of waveguiding performance were carried out using a Monte Carlo ray-tracing method, where photon propagation follows the laws of geometrical optics. Because the plastic scintillator thickness is much larger than the light coherence length, interference effects were neglected. The stochastic nature of the simulations was reflected by not splitting rays at interfaces, but instead treating them as either transmitted or reflected, with probabilities proportional to energy fluxes given by Fresnel's laws. The dependence of these probabilities on the polarization state of the incident ray (e.g., s or p polarization) was also considered. Within the nanocomposite, the inverse transform sampling method was used to generate the optical path length before absorption by the NCs, following an exponential attenuation law determined by the wavelength-dependent absorption cross section, $\sigma(\lambda)$, and the NC concentration, $N(\lambda)$, yielding the attenuation coefficient, $k(\lambda) = \sigma(\lambda)N(\lambda)$. Since the mean path length (inverse of the attenuation coefficient) was always much greater than the average distance between NCs, there was no need to track individual NCs, allowing the nanocomposite (PLMA + NCs) to be treated as a uniform medium. Once absorbed by an NC, the photon's fate – either reemission or nonradiative relaxation – was determined by Monte Carlo sampling based on the experimental emission quantum yield. The direction of reemission was distributed uniformly, and the reemission wavelength was determined using rejection sampling based on the experimentally obtained NC luminescence spectrum. The final fate of each photon was either loss due to nonradiative recombination or escape from the nanocomposite via one of its interfaces. Each simulation typically consisted of $10^5$ - $10^7$ repetitions to achieve adequate statistical averaging. This approach allowed for the evaluation of various observables and the addition of additional processes. The nanocomposite was modelled as a rectangular parallelepiped with dimensions 6.0×0.5×0.1 cm. The experimental illumination condition was modelled by positioning the primary photon origin close to one end of the composite (along the longer axis), while uniform illumination was achieved by randomly generating primary photons within the entire nanocomposite volume. Simulated guided PL spectra and corresponding intensities were reconstructed from the photons escaping from the opposite 0.5×0.1 cm face.

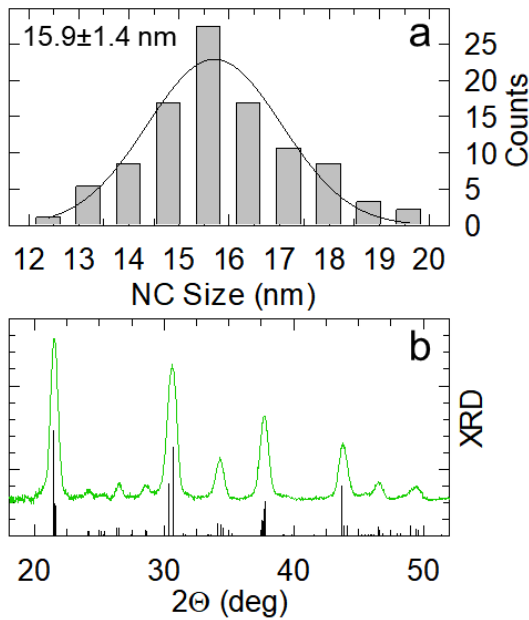

**Supplementary Figure 1: a)** CsPbBr$_3$ NCs size distribution extracted from TEM image in Figure 1 showing an average NC size of 15.9 ± 1.4 nm. **b)** X-ray diffractogram of CsPbBr$_3$ NCs together with the reference (ICSD number 97851) pattern of an orthorhombic CsPbBr$_3$ crystal.

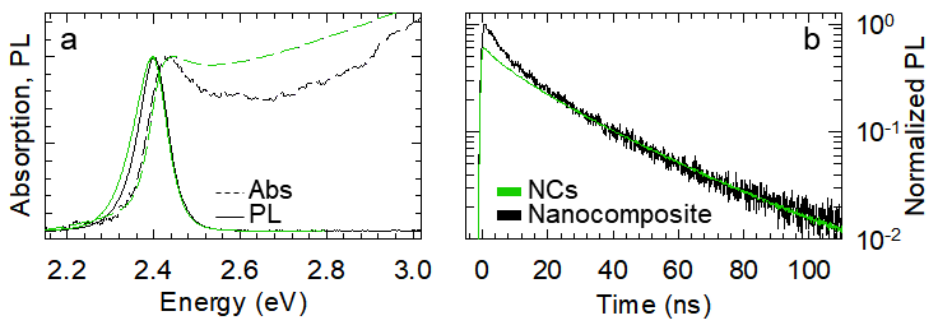

**Supplementary Figure 2: a)** Normalized optical absorption (dashed lines) and PL spectra (solid lines) of CsPbBr$_3$ NCs dispersed in colloidal solution (green) and in a 0.1 wt% loaded PLMA nanocomposite. **b)** Tail normalized PL decay traces of the colloidal solution and the nanocomposite. The faster initial component in the nanocomposite reflects the minor optical losses (~20%) introduced by the polymerization process.

Supplementary References